\PassOptionsToPackage{unicode}{hyperref}
\PassOptionsToPackage{hyphens}{url}

\documentclass[10pt, a4paper]{article}
\usepackage[margin=1in]{geometry}
\usepackage{amsmath,amssymb}
\usepackage{lmodern}
\usepackage{iftex}
\ifPDFTeX
  \usepackage[T1]{fontenc}
  \usepackage[utf8]{inputenc}
  \usepackage{textcomp} 
\else
  \usepackage{unicode-math}
  \defaultfontfeatures{Scale=MatchLowercase}
  \defaultfontfeatures[\rmfamily]{Ligatures=TeX,Scale=1}
\fi
    
\IfFileExists{upquote.sty}{\usepackage{upquote}}{}
\IfFileExists{microtype.sty}{
  \usepackage[]{microtype}
  \UseMicrotypeSet[protrusion]{basicmath} 
}{}
\makeatletter
\@ifundefined{KOMAClassName}{
  \IfFileExists{parskip.sty}{%
    \usepackage{parskip}
  }{
    \setlength{\parindent}{0pt}
    \setlength{\parskip}{6pt plus 2pt minus 1pt}}
}{
  \KOMAoptions{parskip=half}}
\makeatother
\usepackage{xcolor}
\IfFileExists{xurl.sty}{\usepackage{xurl}}{} 
\IfFileExists{bookmark.sty}{\usepackage{bookmark}}{\usepackage{hyperref}}
\hypersetup{
  hidelinks,
  pdfcreator={LaTeX via pandoc}}
\usepackage{longtable,booktabs,array}
\usepackage{multirow}
\usepackage{calc} 
\usepackage{etoolbox}
\makeatletter
\patchcmd\longtable{\par}{\if@noskipsec\mbox{}\fi\par}{}{}
\makeatother
\IfFileExists{footnotehyper.sty}{\usepackage{footnotehyper}}{\usepackage{footnote}}
\makesavenoteenv{longtable}
\usepackage{graphicx}
\makeatletter
\def\maxwidth{\ifdim\Gin@nat@width>\linewidth\linewidth\else\Gin@nat@width\fi}
\def\maxheight{\ifdim\Gin@nat@height>\textheight\textheight\else\Gin@nat@height\fi}
\makeatother
\setkeys{Gin}{width=\maxwidth,height=\maxheight,keepaspectratio}
\makeatletter
\def\fps@figure{htbp}
\makeatother
\usepackage[normalem]{ulem}
\ifLuaTeX
  \usepackage{selnolig}  
\fi

\author{}
\date{}

\usepackage{makecell} 
\usepackage{multirow} 
\usepackage{siunitx} 
\usepackage{geometry} 

\begin{document}

\hypertarget{optimal-design-of-spur-gears-using-particle-swarm-optimization}{%
\section{Optimal Design of Spur Gears using Particle Swarm
Optimization}\label{optimal-design-of-spur-gears-using-particle-swarm-optimization}}

Ricardo Fitas * (1), Carlos Fernandes (2), Carlos Conceição António (3)

\begin{enumerate}
\def\labelenumi{(\arabic{enumi})}
\item 
  Institute of Paper Technology and Mechanical Process Engineering, Technical University of Darmstadt, Darmstadt, Germany
\item
  INEGI/LAETA, Faculty of Engineering, University of Porto, Porto,
  Portugal
\item
  FEUP, Universidade do Porto, Rua Dr. Roberto Frias s/n, 4200-465
  Porto, Portugal
\end{enumerate}

*Email: \href{mailto:ricardo.fitas@tu-darmstadt.de}{\nolinkurl{ricardo.fitas@tu-darmstadt.de}}

\textbf{Abstract}: Optimizing the design of spur gears, regarding their
mass or failure reduction, leads to reduced costs. The proposed work is
aimed at using Particle Swarm Optimization (PSO) to solve single and
multiple-objective optimization problems concerning spur gears. pinion
number of teeth, the module, the face width, and the profile shift
coefficients of both the pinion and wheel. Mass, gear loss factor, specific
sliding, contact ratio, and safety factors are variables considered for
the formulation of the objective function. The results show that those
variables are reduced when compared to the results of the literature.
Mass reduction, for instance, was possible to be achieved due to the
reduction of the gear module and face width, increasing the number of
teeth to achieve the required working center distance of the gear.

\textbf{Keywords}: Particle Swarm Optimization (PSO), Spur Gears, Gear
mass reduction, multi-objective optimization

\begin{enumerate}
\def\labelenumi{\arabic{enumi}.}
\item
  \textbf{Introduction}
\end{enumerate}

The efficiency of gears has been carrying a great focus due to the
existing concerns related to the reduction of carbon dioxide [\ref{ref:1}],
market competition, quality, and regulation laws [\ref{ref:2}]. Gears are
essential to machine elements in transmission systems, and their
efficiency has been linked to reduced costs, increased life [\ref{ref:3}], and
reduced failures [\ref{ref:4}].

Increasing the efficiency of gears leads to optimization problems.
Optimization of gears has been a frequent topic of interest in the
literature [\ref{ref:5}]--[\ref{ref:8}]. From recent publications, various
algorithms and different objectives have been conducted in several
investigations. Miler et al. [\ref{ref:9}] tested the influence of the profile
shift on the problem of the minimization of the gear weight. Results
have shown the increase of the module value and profile shift
coefficients reduce the volume of the gears. Moreover, profile shift
coefficients are shown to be more important than using solely a balance
between gear face width and gear module. Conclusions remark the greatest
influence of the profile shift, which proves the gap in some previous
studies, see, e.g. [\ref{ref:10}] about the non-existence of profile shift
coefficients. Korta and Mundo [\ref{ref:11}] aimed to model a population-based
meta-heuristic approach for optimizing teeth profile on
micro-geometries, using Response Surface Methodology (RSM) for the
quickness of the function evaluation. Daoudi and Boudi [\ref{ref:12}] solved
the minimization problem of the gear weight for four different
materials, using a genetic algorithm. Using multi-objective
optimization, the authors in ref. [\ref{ref:8}] experimented, considering
volume and efficiency as objectives. The results show that the power
loss is lower with higher gear module values. Moreover, profile shift
coefficient values for the wheel, when combined with those of the pinion
and all the experiments, converge towards a higher number of teeth, even
when optimizing the volume.

Hofstetter et al. [\ref{ref:13}] exposed a gearbox design optimization
approach, where load, lifetime, and package requirements are given. After
applying the differential-evolution process, adequate design parameters
are given for the optimized efficiency, package, and cost. The results
show that bearings influence efficiency. Deep groove ball bearings
lead to more comprehensive solutions but are cost-effective, and both
cylindrical and taper roller bearings are, in the opposite way, more
costly but more compact solutions. Other materials can be used to extend
the present study in order to obtain more promising results. In the
following year, the authors in ref. [\ref{ref:3}] applied the non-dominated
sorting genetic algorithm method (NSGA-II) [\ref{ref:14}], which incorporates
fast elitism and removes sharing parameters of genetic algorithm
operators. A decision is made regarding a number of options: Shannon
Entropy, Linear programming technique for multidimensional analysis of
preference, and technique of order preference by similarity to an ideal
solution. The gear module, teeth number, and transmission ratio are
considered design variables, and center distance, bearing capacity, and
meshing efficiency are the design objectives. In 2020, Atila et al.
[\ref{ref:15}] compared five known metaheuristics that solve two different
weighting minimization problems: one introduced in ref. [\ref{ref:16}] used
the width of the gear, the diameters of both the pinion and wheel, the number of teeth of the pinion, and the gear module as the design parameters.

The other problem is an adaptation of the first one by Savsani et al.
[\ref{ref:17}], all of those parameters add to the hardness of the
material the gears are made of is given. Artificial algae algorithm and
Artificial Bee Colony (ABC) are methods that have minimized the
gear weight; Particle Swarm Optimization (PSO) technique is very close
to the results of the two previous ones, but its computational cost is
considerably lower when compared to the other two. In the same year,
Ebenezer et al. [\ref{ref:18}] compared different nature-inspired and Teaching
Learning Based Optimization (TLBO) methods to find the minimum gear
weight, including PSO and Cuckoo Search (CS). These two were found to be
the methods in which performance was the best; the weight is minimized
by 11.41\% when comparing to the gear without profile shift; computation
time of Simulated Annealing (SA) is shown to be lower than both CS and
PSO. Moreover, in 2021, Guilbault and Lalonde [\ref{ref:19}] applied PSO and a
modified Firefly algorithm to minimize the number of initiation points
due to cracks appearing along the use of spur gears, by means of a
fatigue point of view. Two objectives were set: the optimization of
fatigue life concerning the reduction of dynamic loads and dynamic
transmission errors (along a set of pinion speeds); the testing of the
influence of certain variables as the starting point, the correction
amount, and the curvature radius on the optimization of the dynamic
transmission errors and dynamic factors. In the same year, Tomori
[\ref{ref:20}] used a sample gear pair to optimize the choice of the profile
shift coefficients according to different criteria. Tavčar et al.
[\ref{ref:21}] aimed to optimize polymer gears using center distance, cost,
life span, transmission ratio, torque, and speed as input values, and,
after material selection, it optimizes tooth root stress, flank
pressure, flash temperature, tooth wear and deformation, cost, and
volume.

Still, the exposed publications frequently look to minimize the
weight of both the pinion and wheel or the distance between their respective
centers. Increasing the expected output power and efficiency are
other considered objectives, and there are others that are not
frequently used. Tomori [\ref{ref:20}] uses other possible objectives as
methods for the computation of the correspondent profile shift
coefficients, such as the minimum friction loss on the tooth profile,
the optimal bending stress at the root of the pinion, optimal Hertzian
stress, optimal Almen product, the optimal magnitude of the resulting
flash temperature, optimal lubricant film thickness, or optimal linear
wear. However, it does not allow the variation of other gear design
parameters. Also, Atila et al. [\ref{ref:15}] state that control parameters
are difficult to set up onto gear design problems. However, convergence
is mainly not shown or explained, and parameters have not varied in
recent publications. Moreover, Tavčar et al. [\ref{ref:21}] refer that the
the resulting framework has a lack of consideration factors such as tooth
profile modifications, thermal response, and the influence of fillers on
gearing tribology; [\ref{ref:3}] makes also reference to the lack of
multi-optimization solutions; Miler et al. [\ref{ref:9}] has referenced to the
optimization of the specific sliding and its use as a constraint, due to
gearing longevity, which was then studied by Rai and Barman [\ref{ref:22}] in
helical gears.

The proposed work aims to use the PSO technique to study a variety of
multi-optimization problems on spur gears resulting from the combination
of the following objectives: gear loss factor (HVL), contact ratio,
specific sliding, flank pressure safety factor, root stress safety
factor and weight. Increasing contact ratio and safety factors decrease
mechanical failure risk due to bending and contact stresses [\ref{ref:23}],
problems in which adequate gear designs and lubrication are demanded
[\ref{ref:20}]. Decreasing the maximum specific sliding reduces the risk of
sudden breakage of the lubrication film, metal contact, and seizure
damage. Also, increasing efficiency leads to energy savings due to the
lower power supply to obtain the same output power. The gear mass
reduction is carried out in some of the literature since a lighter gear
prevents dealing with higher costs; therefore, the gear mass will be
also considered as an optimization objective. The adequate spur gear is
expected to accomplish most of the objectives. The design variables are
the pinion number of teeth, the module, the face width, and the profile
shift coefficients of both the pinion and wheel. Also, it aims to
investigate and compare different PSO variants and population sizes to
obtain the most suitable metaheuristic configuration.

From previous results from the literature [\ref{ref:5}], it can be
hypothesized that, for the same center distance, the profile shift
coefficients and the face width values may change considerably since
objectives such as the specific sliding, safety factors, and contact
the ratio is added. In contrast, the number of teeth and modules will remain
approximately the same to maintain the compromise between the high
safety factors and the remaining objectives. A literature review
[\ref{ref:24}] indicated that lower values for the module, when combined with
the use of higher values for the number of teeth resulted in the
minimization of weight, or using higher values for the module but lower
values for the face width. It is also expected, for the considered
transmission ratio and center distance, difficulty in finding more gear
designs since the set of the possible number of teeth for the pinion is
lower, resulting in a more limited set of possible designs.

This paper is structured as follows: the methodology section presents
the adopted framework on which the numerical search is based; then,
numerical results are presented, as well as all the constraints and
constant values, e.g., of the material properties; conclusions have to
The most important points from the study carried out in this work are as follows:

\begin{enumerate}
\def\labelenumi{\arabic{enumi}.}
\setcounter{enumi}{1}
\item
  \textbf{Framework concepts}

  \begin{enumerate}
  \def\labelenumii{\arabic{enumii}.}
  \item
    \emph{Multi-objective optimization}
  \end{enumerate}
\end{enumerate}

Let \(N\) be the number of design variables, defined by the
\(\mathbf{x} = \left( x_{1},\ldots,\ x_{N} \right)\) and
\(f(\mathbf{x})\) an objective function. The number of inequality
constraints is given by \(n_{g}\) and the functional inequality
constraints are \(g_{i}\left( \mathbf{x} \right),\ i = 1,\ldots,n_{g}\).
Moreover, the number of equality constraints is given by \(n_{h}\) and
\(h_{j}\left( \mathbf{x} \right),\ j = 1,\ldots,n_{h}\), are the
functional equality constraints. \(\mathbf{x} \in \ S^{N}\), where
\(S^{N} = \left\lbrack x_{1_{L}},x_{1_{U}} \right\rbrack\  \times \ldots \times \left\lbrack x_{N_{L}},x_{N_{U}} \right\rbrack\ \)
is the search space. Each interval in this Cartesian product is
associated to the side constraints of the optimization problem
formulation. Moreover, let there have \(M\) objectives, one may be
referred to as multi-objective optimization problems, given as follows:
\begin{equation*}
\begin{aligned}
& \text{Minimize:} & & \mathbf{F}(\mathbf{x}) = (f_1,...,f_M) \\
& \text{Subject to:} & & g_i(\mathbf{x}) \leq 0, \; i = 1, ..., n_g \\
& & & h_j(\mathbf{x}) = 0, \; j = 1, ..., n_h \\
& & & x_{k,\text{min}} \leq x_k \leq x_{k,\text{max}}, \; k = 1, ..., N
\end{aligned}
\end{equation*}

According to Gunantara [\ref{ref:24}] and António [\ref{ref:25}], two main methods
divide the way one deals with multi-objective optimization: Pareto
dominance and scalarization. In Pareto dominance, a set of solutions
called the Pareto set is defined as solutions that dominate all the
other solutions within the search space, i.e., the solutions that are
not dominated. Therefore, no element of the Pareto set can dominate any
of the other candidate elements. Dominance can be defined by the
mathematical statement in the next equation, where
\(\mathbf{x}_{\mathbf{1}},\ \mathbf{x}_{\mathbf{2}} \in S^{N}\)
[\ref{ref:26}].

\begin{align*}
\mathbf{x}_{\mathbf{1}} \prec \mathbf{x}_{\mathbf{2}} \; \text{:} \; \forall i = 1, ..., M, f_i(\mathbf{x}_{\mathbf{1}}) \leq f_i(\mathbf{x}_{\mathbf{2}}) \; \land \; \exists i \in \{1, ..., m\}, f_i(\mathbf{x}_{\mathbf{1}}) < f_i(\mathbf{x}_{\mathbf{2}})
\end{align*}

The Pareto set is a set with multiple valid solutions. An adequate
solution is left for the user to choose after optimization. In
scalarization, the various objectives are combined in a single fitness
function to transform the problem into a single optimization problem
[\ref{ref:25}].

The current approach, explained in Section 3, only considers the
scalarization method.

\begin{enumerate}
\def\labelenumi{\arabic{enumi}.}
\setcounter{enumi}{1}
\item
  \emph{Particle Swarm Optimization}
\end{enumerate}

\begin{figure}[ht]
\centering
\includegraphics[width=\linewidth]{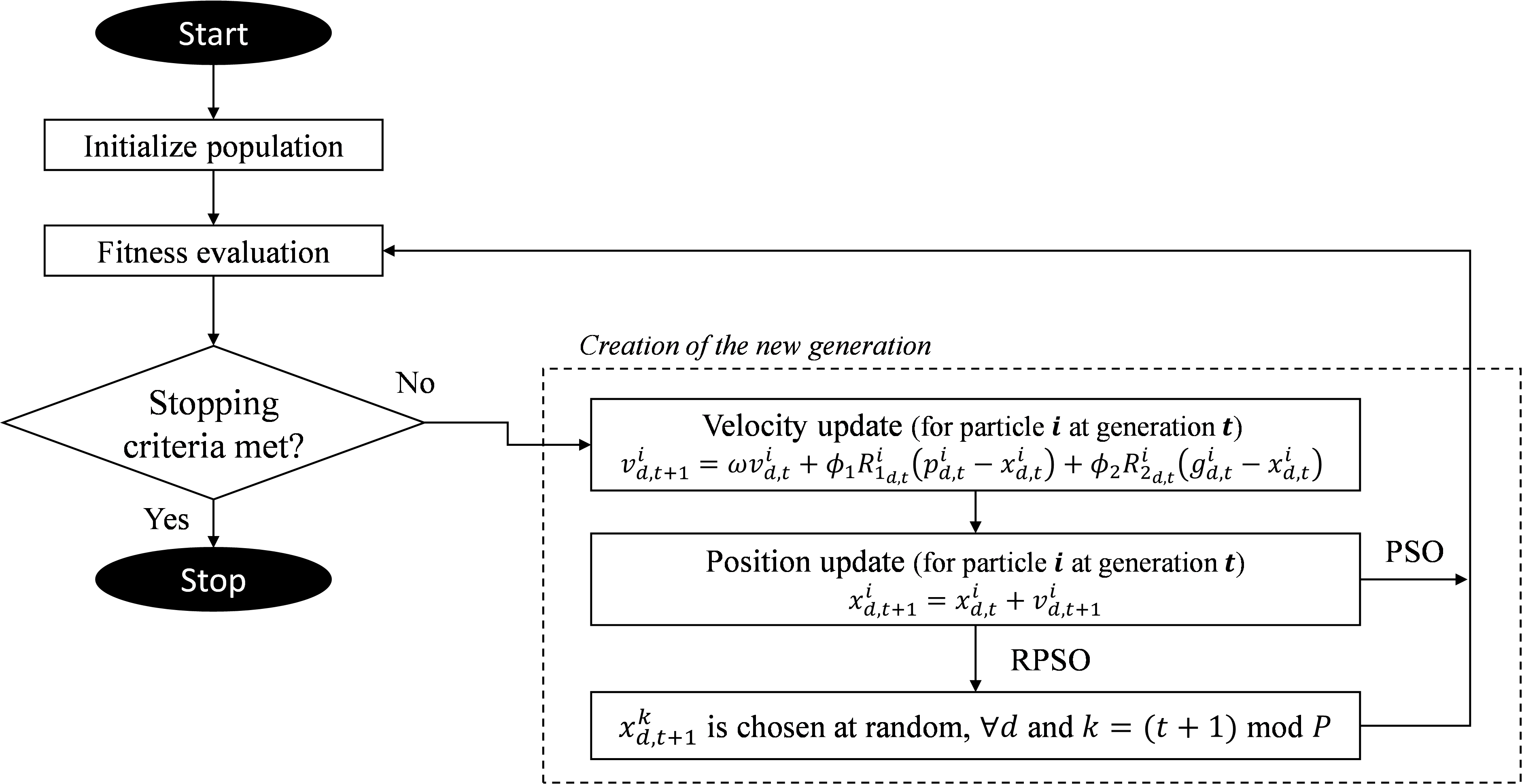}
\caption{Flowchart for the Classic PSO Algorithm / proposed RPSO Algorithm.}
\label{fig:1}
\end{figure}

PSO is a bio-inspired population-based technique with plenty of
applications in engineering optimization problems. It was inspired by
social behaviors like bird flocking and fish schooling [\ref{ref:27}],
[\ref{ref:28}]. PSO has a generalized flowchart represented in Figure \ref{fig:1}.
After the generation of the starting population, which is usually done
at random, the fitness of the solutions, also called particles, is
evaluated with respect to the problem it aims to optimize.
Therefore, the particles' velocity and positions are updated based on
mathematical expressions and dependently on the fitness values of the
particles. Velocities and positions of each population particle \(i\),
at each generation \(t\), and for each dimension \(d\), as it is shown
as follows:

\begin{align}
v_{d,t+1}^{i} &= \omega v_{d,t}^{i} + \phi_1 R_{1d,t}^{i} (p_{d,t}^{i} - x_{d,t}^{i}) + \phi_2 R_{2d,t}^{i} (g_{d,t}^{i} - x_{d,t}^{i}) \\
x_{d,t+1}^{i} &= x_{d,t}^{i} + v_{d,t+1}^{i}
\end{align}

In the equations above, \(v\) is the velocity of a particle,
\(\mathbf{x}\) is the position of a particle, \(\omega\) is the inertia
weight coefficient, which has then been introduced in later publications
[\ref{ref:29}], \(\phi_{1}\) and \(\phi_{2}\) are cognitive and social
coefficients, respectively, \(\mathbf{p}_{t}^{i}\) is the best position
of a particle \(i\) among all the generations to the generation \(t\),
\(\mathbf{g}\) is the best position among all particles and generations
to the generation \(t\), and \(R_{1}\) and \(R_{2}\) are values sampled
at random.

PSO was first implemented in 1995. This has led the scientific
community to verify some issues, resulting in the most known premature
convergence and consequently developing variations of PSO. In 1999,
Clerc [\ref{ref:30}] has introduced a constriction factor related to
\(\phi_{1}\) and \(\phi_{2}\). In [\ref{ref:31}], discrete-time PSO is
generalized to a continuous-time PSO from an analytical point of view.
The constriction factor is multiplied by the right-hand side of (1).
Random PSO, abbreviated as RPSO [\ref{ref:32}], is another PSO variation that
is aimed at finding the global minimum. For that, a particle is chosen and
sampled at random. In 2006, Van der Bergh [\ref{ref:32}] demonstrated that
PSO is a local but not a global search algorithm. However, RPSO has been
proven to be a global search method. Since particles are sampled
randomly into the search space, the final solution is not dependent on
the starting population. The same RPSO algorithm that has been used in
different applications [\ref{ref:33}]--[\ref{ref:35}] is now used for the
optimization of spur gears.

\begin{enumerate}
\def\labelenumi{\arabic{enumi}.}
\setcounter{enumi}{2}
\item
  \textbf{Methodology}
\end{enumerate}

The proposed work consists of optimizing gear design parameters, given
the minimization function, \(F\), and inequality and equality
constraints that will be defined later on. This work considers that the
transmission ratio and the working center distance, \(u\) and \(a’\),
respectively, are given, which leads to choosing between multiple gears
that respect the given \(u\) and \(a’\). Due to the natural
discretization of the number of teeth of both pinion and wheel,
\(z_{1}\) and \(z_{2}\), respectively, one may not find suitable or even
possible gear combinations, leading to the consideration of an
admissible slight variation of the transmission ratio. That variation
shall be not higher than a given tolerance \(\text{to}l_{u}\). In its
turn, \(a’\) does not come with any possible tolerance. The total number
of inputs is given in Table \ref{tab:1}.

\begin{longtable}[]{@{}
  >{\raggedright\arraybackslash}p{(\columnwidth - 2\tabcolsep) * \real{0.5000}}
  >{\raggedright\arraybackslash}p{(\columnwidth - 2\tabcolsep) * \real{0.5000}}@{}}
 \caption{Requirement parameters for the possible gears to consider} \label{tab:1} \\
\toprule
\begin{minipage}[b]{\linewidth}\raggedright
Parameter
\end{minipage} & \begin{minipage}[b]{\linewidth}\raggedright
Definition
\end{minipage} \\
\midrule
\endhead
\(u\) & Transmission ratio \\
\(\text{to}l_{u}\) & Tolerance of the transmission ratio \\
\(a’\) & Working axis distance \\
\bottomrule

\end{longtable}

The transmission ratio is expressed as follows:

\begin{equation}
u = \frac{z_2}{z_1}
\end{equation}
where \(z_{1}\) and \(z_{2}\) are the number of teeth for the pinion and
the wheel, respectively. Depending on \(\text{tol}_{u}\), \(z_{2}\) can
be a direct function of \(z_{1}\) or not. This dependency is, in turn,
dependent on the value of \(\text{tol}_{u}\), being \(z_{2}\) value
restricted according to eq. (4).

\begin{equation}
tol_u \geq \left| \frac{z_2}{z_1} - u \right|
\end{equation}

Here, \(z_{1}\) is also restricted by \(\text{tol}_{u}\), as it is shown
in eq. (5) since it is necessary for at least one possible \(z_{2}\).
Note that \(\left\lfloor \left. \  \cdot \right\rceil \right.\ \) in (5)
refers to the nearest integer notation.

\begin{equation}
tol_u \geq \left\lfloor{ \frac{\lfloor z_1 \cdot u \rceil}{z_1} - u }\right\rceil
\end{equation}

The axis distance without any profile shift is calculated using eq. (6).

\begin{equation}
a = r_1 + r_2 = \frac{z_{1}m}{2} + \frac{z_{2}m}{2} = a(z_1, z_2, m)
\end{equation}

where \(r_{1}\)and \(r_{2}\) are the pitch radii, respectively, and
\(m\) is the profile module, which is equal for both pinion and wheel.
The working axis distance is dependent on gear standard parameters and
the optimized profile shifts. Two conditions may be necessary to know
this incognita: the base pitch is the same whichever the profile shift
is, and the pitch is the sum of the thicknesses of the pinion and wheel when
considering the working conditions, as according to eq. (7).

\begin{equation}
\begin{cases}
\text{inv} \alpha' = \text{inv} \alpha + 2 \cdot \tan \alpha \cdot \frac{\Sigma x}{z_1 + z_2} \\
a \cdot \cos \alpha = a' \cdot \cos \alpha' \\
\Sigma x = x_1 + x_2
\end{cases}
\end{equation}

In eq. (7), the involute function \(\text{inv\ }(x) = \tan(x) - x\)
gives the involute of the angle \(x\). The function dependence
represented in eq. (8) must be in the count to accomplish the initial
constraint.

\begin{equation}
\alpha' = \alpha' \left( \Sigma x, \alpha, z_1, z_2, a(z_1, z_2, m) \right) = \alpha' \left( \Sigma x, \alpha, z_1, z_2, m \right)
\end{equation}

Eq. (8) remarks an important conclusion: \(a’\) will be not dependent of
\(x_{1}\) or \(x_{2}\) individually, since only their sum is explicitly
written in eq. (7).

Therefore, and because \(a’\) is given, the sum of the profile shifts
may be turned on the only unknown of eq. (7), being \(x_{2}\) the only
parameter to be calculated from the rest of the gear parameters.

The profile shift is considered in this work to accomplish the axis
distance value; still, this is typically done to accomplish the equality
sliding speeds, which is also intended to minimize in this work.

There are many constraints that may be considered due to the
real-scenario application. Firstly, \(\alpha\) and \(m\) are normalised
values. The module is one of the possible values from any of the listed
sets \(\mathbf{L}_{\mathbf{A}}\) and \(\mathbf{L}_{\mathbf{B}}\)
(according to DIN 780 Part 1), being \(\mathbf{L}_{\mathbf{A}}\) the
most preferable.

\begin{align*}
\mathbf{L}_{\mathbf{A}} = \{& 0.05,\ 0.06,\ 0.08,\ 0.1,\ 0.12,\ 0.16,\ 0.2,\ 0.25,\ 0.3,\ 0.4,\ 0.5,\ 0.6,\ 0.7,\ 0.8,\\ & 0.9,\ 1,\ 1.25,\ 1.5,\ 2,\ 2.5,\ 3,\ 4,\ 5,\ 6,\ 8,\ 10,\ 12,\ 16,\ 20,\ 25,\ 32,\ 40,\ 50,\ 60\}\\
\mathbf{L}_{\mathbf{B}} = \{& 0.055,\ 0.07,\ 0.09,\ 0.11,\ 0.14,\ 0.18,\ 0.22,\ 0.28,\ 0.35,\ 0.45,\ 0.55,\ 0.65,\ 0.75,\ 0.85,\\ & 0.95, \ 1.125,\ 1.375,\ 1.75,\ 2.25,\ 2.75,\ 3.25,\ 3.5,\ 3.75,\ 4.25,\ 4.5,\ 4.75,\ 5.25,\ 5.5,\ 5.75,\\ & 6.5,\ 7,\ 9,\ 11,\ 14,\ 18,\ 22,\ 27,\ 28,\ 30,\ 36,\ 39,\ 42,\ 45,\ 55,\ 70\}
\end{align*}

Also, the values of the profile shifts are restricted, as in eq. (9).

\begin{equation}
    \begin{aligned}
        S_{\text{min}} &\leq \sum_{i=1}^2 x_i \leq S_{\text{max}} \\
        x_{1,min} &\leq x_1 \leq x_{1,max} \\
        x_{2,min} &\leq x_2 \leq x_{2,max}
    \end{aligned}
\end{equation}

where \(S_{\min} = - 1\), \(S_{\max} = 2\), \(x_{1,min} = 0,\)
\(x_{1,max} = 1\), \(x_{2,min} = - 0.5,\) \(x_{2,max} = 1\) for this
work. The minimum profile shifts for pinion and wheel are limited by the
non-occurrence of undercut, written in eq. (10).

\begin{equation}
x_{i_{\text{min}}} = \frac{z' - z_i}{z'} \quad i = 1,2
\end{equation}

where \(z^{'} = 2{(\sin\alpha)}^{- 2}\). Also,
\(x_{i,\ min}^{*},\ \ i = 1,2\) are restricted according to eq. (11).

\begin{equation}
x_{i,\text{min}} =
\begin{cases}
0, & \text{if } i = 1 \land x_{i,\text{min}} < 0 \\
x_{i,\text{min}}, & \text{otherwise}
\end{cases}
\end{equation}

\begin{figure}
\centering
\includegraphics[width=0.85\linewidth]{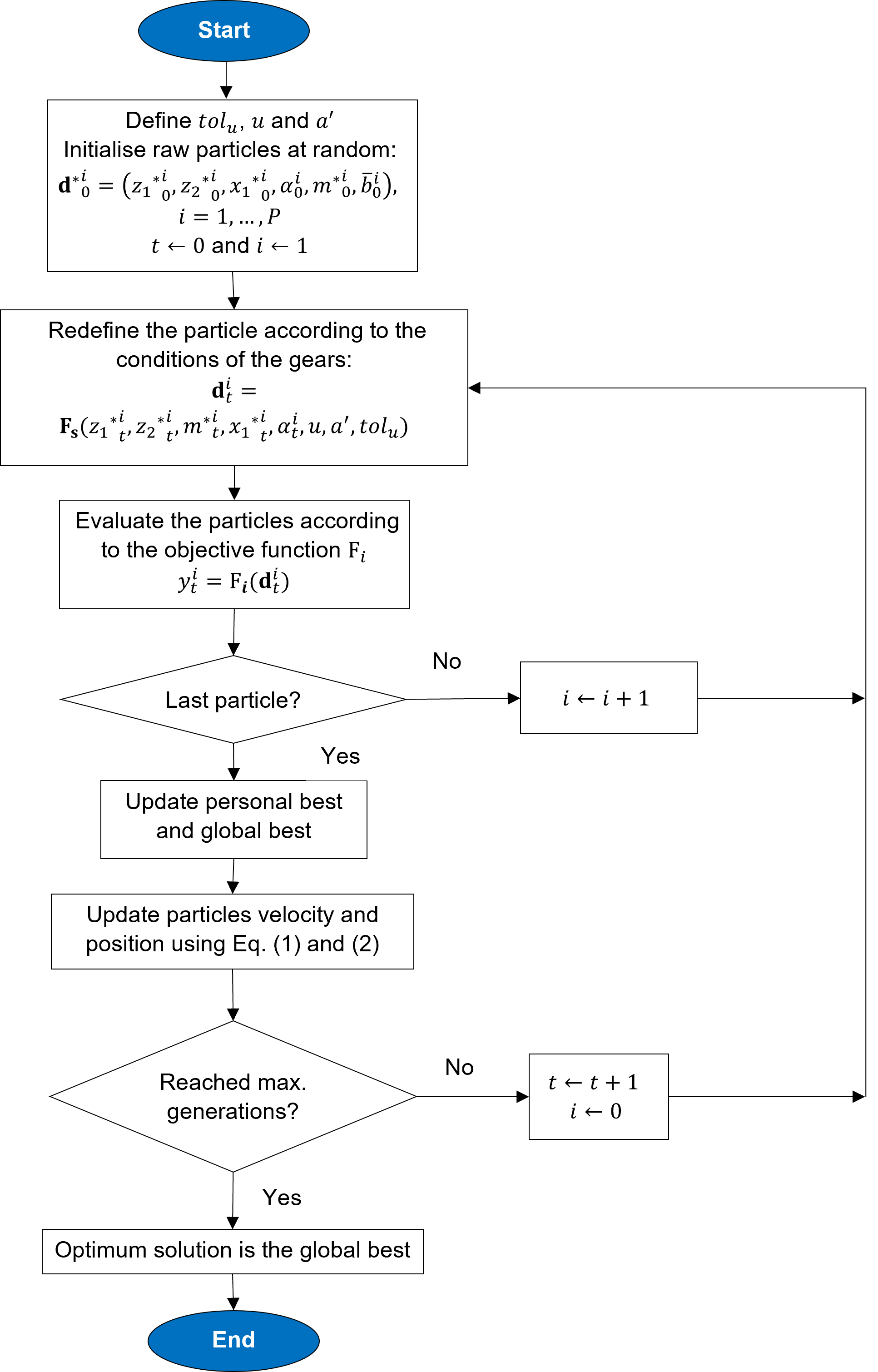}
\caption{Proposed flowchart for the optimization process.}
\label{fig:2}
\end{figure}

Using the previous equations, a proposed algorithm presented in Figure \ref{fig:2}
is aimed to search for a valid gear. Given the variables
\({z_{1}}^{*},{z_{2}}^{*},m^{*},{x_{1}}^{*},\alpha\), \(\overline{b}\),
where \(\overline{b}\) is the normalized face width of the gear, the
calculation of a valid gear is done based on a function called
\(\mathbf{F}_{\mathbf{s}}\).This function is then aimed at transforming
a given set of initial parameters into another set that accomplishes the
aforementioned constraints of center distance and transmission ratio.
The output of the proposed algorithm corresponds to the vector of the design
variables
\(\mathbf{d} = \left( z_{1},z_{2},x_{1},x_{2},m,\alpha,b \right)\). In
the algorithm, \(z_{1},z_{2},x_{1}\) and \(m\) are modified around a
fixed starting point \(({z_{1}}^{*},{z_{2}}^{*},m^{*},{x_{1}}^{*})\),
and the rest of the missing parameters, i.e., \(x_{2}\) and
\(b = \ 5m\left( \overline{b} + 1 \right)\), where \(b\) is the face
width of the gear, are directly calculated once it is known that the
resulting gear is compatible with the given inputs for the transmission
ratio, its tolerance, and the working axis distance.

The scalar functions that will be used later to defined multi-objective
functions are now defined. These are the contact ratio (\(f_{1}\)), the
mass of the gear (\(f_{2}\)), the root safety factor (\(f_{3}\)), the
flank safety factor (\(f_{4}\)), the gear loss factor (\(f_{5}\)) and
the difference between specific sliding (\(f_{6}\)).

Considering \(r_{\text{bi}} = r_{i}\ \cos\alpha\ \)and
\(r_{\text{ai}}^{'} = r_{i} + m\left( 1 + x_{i} \right),\ i = 1,2\), the
contact ratio is defined as follows:

\begin{equation}
f_1 = f_1(\mathbf{d}) = \varepsilon_{\alpha} = \frac{\sqrt{r_{a1}^2 - r_{b1}^2} + \sqrt{r_{a2}^2 - r_{b2}^2 - a' \cdot \sin \alpha'}}{p \cdot \cos \alpha}
\end{equation}

The adopted mass used in the present work is according to eq. (13). In
the equation, \(b\) is the face width of the gear, \(\rho\) is the specific gravity of the material of the gear,
\({r_{a}}_{1} = r_{1} + x_{1}m\) and \({r_{a}}_{2} = r_{2} + x_{2}m\).

\begin{equation}
f_2 = f_2(\mathbf{d}) = W = p\pi b\left( r_{a1}^2 + r_{a2}^2 \right)
\end{equation}

The root (\(S_{H}\)) and flank (\(S_{F}\)) safety factors are
accordingly to eq. (14) and eq. (15), respectively.

\begin{equation}
f_3 = f_3(\mathbf{d}) = S_H = \frac{\sigma_{H_a}}{\sigma_H}
\end{equation}

\begin{equation}
f_4 = f_4(\mathbf{d}) = S_F = \frac{\sigma_{F_a}}{\sigma_F}
\end{equation}

The contact pressure \(\sigma_{H}\) is defined:

\begin{equation}
\sigma_H = Z_{D} \sigma_{H0} \sqrt{\prod K_i}
\end{equation}

\begin{equation}
\sigma_{H0} = \sqrt{\frac{F_t}{2r_{1}b} \frac{u + 1}{u}} \prod Z_i 
\end{equation}

where \(Z_{D}\) is the pinion single pair tooth contact factor,
\(Z_{i}\) are other contact factors, such as the zone factor, the
contact ratio factor and the elasticity factor, \(K_{i}\) are correction
factors, such as the application factor, the dynamic factor and the face
and the transverse load factors for contact stress, and \(F_{t}\) is the
tangential load. \({\sigma_{H}}_{a}\) is defined as the allowed contact
pressure. Moreover, the flank pressure \(\sigma_{F}\) is defined as
follows:

\begin{equation}
\sigma_F = \frac{F_t}{bm} \prod Y_i \prod K_i
\end{equation}

where \(Y_{i}\) and \(K_{i}\) are correction factors that depend on
system characteristics such as the lubricant, roughness, or the
numerical method being used for their calculation. \({\sigma_{F}}_{a}\)
is defined as the allowed flank pressure. The gear loss factor
\(H_{\text{VL}}\ \)is defined as follows:

\begin{equation}
f_5 = f_5(\mathbf{d}) = H_{VL} = \frac{1}{p_b}\int_{A}^{B} \frac{F_N(x)v_g(x)}{F_{bt}v_{tb}} dx
\end{equation}

In (19), \(p_{b} = \pi m\cos\alpha\) is the base pitch, A and B are the
gearing starting and ending points, respectively, \(v_{g}(x)\) is the
sliding speed on a point \(x\), \(F_{N}(x)\) is the normal load in
relation to the base radius, \(F_{\text{bt}}\) is the tangential load in
relation to the base radius and \(v_{\text{tb}}\) is the tangential
component of the sliding speed in relation to the base radius. The
difference of the specific sliding speeds \(\Delta g\) is given as
follows:

\begin{equation}
f_6 = f_6(\mathbf{d}) = \Delta g = \left| \left| 1- \frac{z_1}{z_2} \cdot \frac{\sqrt{r_{a2}^{' 2} - r_{b2}^2}}{a \cdot \sin \alpha - \sqrt{r_{a2}^2 - r_{b2}^2}} \right| - \left|\frac{z_2}{z_1} \cdot \frac{\sqrt{r_{a1}^{' 2} - r_{b1}^2}}{a \cdot \sin \alpha - \sqrt{r_{a1}^2 - r_{b1}^2}} - 1 \right| \right|
\end{equation}

The functional constraints are represented in eq. (21) and eq. (22).

\begin{equation}
    \begin{aligned}
    g_1(d) &= tol_{u}(d) - tol_{u}^M \\
    g_2(d) &= \varepsilon_{\alpha}^M - \varepsilon_{\alpha}(d) \\
    g_3(d) &= S_H^M - S_H(d) \\
    g_4(d) &= S_F^M - S_F(d)
    \end{aligned}
\end{equation}

\begin{equation}
    \begin{aligned}
        h_1(d) &= \alpha'(d) - \alpha'_g \\
        h_2 &= u - u_g
    \end{aligned}
\end{equation}

In the current work, \(\epsilon_{\alpha}^{M} = 1.45\),
\(S_{H}^{M} = 1.4\) and \(S_{F}^{M} = 2.0\) are the allowed minimum
contact ratio, minimum root safety factor and minimum flank safety
factor, respectively. The maximum transmission tolerance ratio
\(\text{to}l_{u}^{M}\), the exact working axis distance \(a_{e}^{'}\)
and the desired transmission ratio \(u_{e}\) are defined for each case
study later on.

\addtocounter{table}{-1}
\begin{longtable}[]{@{}
  >{\raggedright\arraybackslash}p{(\columnwidth - 0\tabcolsep) * \real{1.0000}}@{}}
\toprule
\begin{minipage}[b]{\linewidth}\raggedright
\textbf{Algorithm 1} Gear Search Algorithm (\(\mathbf{F}_{\mathbf{s}})\)
\end{minipage} \\
\midrule
\endhead
\textbf{Given}: \(\alpha,u,a',\text{tol}_{u}\) (Optional variables:
\({z_{1}}^{*},{z_{2}}^{*},m^{*},{x_{1}}^{*},\ \overline{b}\))

\textbf{Initialize}: \({z_{1}}^{*},{z_{2}}^{*},m^{*},{x_{1}}^{*}\),
\(\overline{b}\) (if not previously given)

1:
\({{z_{1}}^{*}}_{0},\ {z_{1}}^{*} \leftarrow \text{int}\left( {z_{1}}^{*} \right);\)
discretize \(m^{*}\) and call it \(m^{*}\) ,\({m^{*}}_{0};\)

2: With \(\alpha\), use (10) to limit \({x_{1}}^{*}\).

3:
\(T_{{z_{1}}^{*}}^{*} \leftarrow \left\{ {m^{*}}_{0} \right\}\text{.\ }\)Evaluate the minimum constant \(k_{{z_{1}}^{*}}\mathbb{\in N:}\) \(z_{2}^{*} = (z_{1}^{*} + k_{z_{1}^{*}}c)(u + \varepsilon \, \text{tol}_{u}) \in \mathbb{N}, \forall c \in \mathbb{Z}^{*} \land \exists \varepsilon \in [-1,1] \subset \mathbb{R},\)where
\(\mathbb{Z}^{*} \subset\) \(\mathbb{Z}\) accordingly to (5).
\(Z \leftarrow \left\{ c \right\}.\)

4: According to the condition defined in Step 3, define a set
\(E_{{z_{1}}^{*}} \text{ of all possible } \varepsilon \text{. If}\)
\(E_{{z_{1}}^{*}} \equiv \varnothing\), go to Step 11.

5: Apply (3) to find
\({{z_{2}}^{*}}_{0},\ {z_{2}}^{*} \leftarrow {z_{1}}^{*}\left( u + {\varepsilon^{*}\text{\ tol}}_{u} \right),\ \varepsilon^{*}:\min{\{|\varepsilon|,\ \varepsilon \in E_{{z_{1}}^{*}}\}}\).
\(E_{{z_{1}}^{*}}^{*} \leftarrow \{\varepsilon^{*}\}\)

6: With \({z_{1}}^{*},{z_{2}}^{*},m^{*},\ \alpha\) and \(z'\) calculate
the limits of \({x_{1}}^{*}\) and \({x_{2}}^{*}\) using (9); use (11).
\(flag \leftarrow 0\)

7: Use (7) to calculate \(\sum_{}^{}x\) ;
\({x_{2}}^{*} \leftarrow \sum_{}^{}x - {x_{1}}^{*}\).

8: If the results on Steps 6 and 7 are according to (9), then
\(z_{1} \leftarrow {z_{1}}^{*},\ z_{2} \leftarrow {z_{2}}^{*},\ m \leftarrow m^{*},\)
\(x_{1} \leftarrow {x_{1}}^{*}\) , \(x_{2} \leftarrow {x_{2}}^{*}\) and
\(b = 5m\left( \overline{b} + 1 \right)\) and exit the function. Else,
and if flag \(\leftarrow 0\), then \({x_{1}}^{*} \leftarrow \text{rand}(0,1)\),
flag \(\leftarrow 1\) and go back to Step 7. Else, go to Step 9.

9: If
\(T_{{z_{1}}^{*}}^{*} \equiv \mathbf{L}_{\mathbf{A}} \cup \mathbf{L}_{\mathbf{B}}\)
go to Step 10. Otherwise,
\(m^{*} \leftarrow \min{\{|m_{T_{{z_{1}}^{*}}} - {m^{*}}_{0}|}\}\),
where
\(T_{{z_{1}}^{*}}\):\(= L_{B}\backslash T_{{z_{1}}^{*}}^{*} \Leftarrow T_{{z_{1}}^{*}}^{*} \subseteq L_{A}\  \land \ L_{A}\backslash T_{{z_{1}}^{*}}^{*} \Leftarrow T_{{z_{1}}^{*}}^{*} \nsubseteq L_{A}\).
\(T_{{z_{1}}^{*}}^{*} \leftarrow \{ m^{*}\} \cup T_{{z_{1}}^{*}}^{*}\).
Go back to Step 6.

10: If \(E_{{z_{1}}^{*}}^{*} \equiv E_{{z_{1}}^{*}}\ \)go to Step 11.
Else,
\({z_{2}}^{*} \leftarrow {z_{1}}^{*} \left( u + {\varepsilon^{*} tol}_{u} \right),\ \varepsilon^{*}:\min\left\{ |\varepsilon|,\ \varepsilon \in E_{{z_{1}}^{*}}\backslash E_{{z_{1}}^{*}}^{*}\not\equiv\varnothing \right\};\ E_{{z_{1}}^{*}}^{*} \leftarrow \{\varepsilon^{*}\} \cup E_{{z_{1}}^{*}}^{*}\)
and go back to Step 6.

11:
\({\ c^{*} \leftarrow \min{\{|{z_{1}}^{*} + k_{{z_{1}}^{*}} c - {{z_{1}}^{*}}_{0}|},\ c \in \mathbb{Z}^{*}\backslash Z\}.\ \ Z \leftarrow \{ c^{*}\} \cup Z\text{.\ \ }z_{1}}^{*} \leftarrow {z_{1}}^{*} + k_{{z_{1}}^{*}} c^{*}\).
\(T_{{z_{1}}^{*}}^{*} \leftarrow \left\{ {m^{*}}_{0} \right\}.\) Go back
to Step 4.

\textbf{Return}: \(z_{1},z_{2},x_{1},x_{2},m,\ \alpha,b\ \) \\
\bottomrule
\end{longtable}

The flowchart that represents the optimization procedure is represented
in Figure \ref{fig:2}. Initially, gear configurations are generated at random as a
standard procedure of PSO. Then, \(\mathbf{F}^{\mathbf{*}}\) is used to
correct the randomized configuration, and that obeys the conditions of
center distance, transmission ratio, and tolerance. After the procedure,
the resulting gears are evaluated, and the best is considered. PSO and
RPSO are used for the search for new potential candidate configurations.
The process repeats, including the evaluation of
\(\mathbf{F}^{\mathbf{*}}\mathbf{\ }\)until the stopping criteria are
met.

Due to the significant number of possible variables that can be
optimized and the number of functions that can be minimized, a total of
three comparative studies of the designed algorithm have been conducted:

\begin{itemize}
\item
  Study 1. variation of the tolerance \(\text{to}l_{u}^{M}\): the
  tolerance related to the transmission ratio variation is set to two
  different values, 0.05 and 0.2.
\item
  Study 2. Variation of the objective function: using
  \(\text{to}l_{u}^{M} = 0.2\), a single-objective function \(F_{1}\)
  and total of three multi-objective functions (\(F_{2},\ F_{3}\) and
  \(F_{4}\) are aggregation functions) have been considered:

  \begin{itemize}
  \item
    \begin{quote}
    \(F_{1}\): Mass minimization (also conducted in studies 1 and 3):
    \end{quote}

    \begin{equation}
    \begin{aligned}
    \text{Minimize:} \quad & F_{1}(\mathbf{d}) = f_2(\mathbf{d}) \\
    \text{Subject to:} \quad & g_i(\mathbf{d}) \leq 0 \\
    & h_k(\mathbf{d}) = 0 \\
    & d_{i,L} \leq d_i \leq d_{i,u}, \; i=1, ..., N_d
    \end{aligned}
    \end{equation}
\item
  \begin{quote}
  \(F_{2}\): Mass minimization and maximization of \(S_{H}\) and
  \(S_{F}\):
  \end{quote}

\begin{equation}
    \begin{aligned}
\text{Minimize:} \quad & F_{2}(\mathbf{d}) = \sqrt{f_2(\mathbf{d})^2 + \left( \frac{1}{1 + f_3(\mathbf{d})} \right)^2 + \left( \frac{1}{1 + f_4(\mathbf{d})} \right)^2} \\
\text{Subject to:} \quad & g_i(\mathbf{d}) \leq 0 \\
& h_k(\mathbf{d}) = 0 \\
& d_{i,L} \leq d_i \leq d_{i,u}, \; i=1, ..., N_d
\end{aligned}
\end{equation}

\item
  \begin{quote}
  \(F_{3}\): Minimization of the gear mass and \(H_{\text{VL}}\);
  \end{quote}

\begin{equation}
    \begin{aligned}
\text{Minimize:} \quad & F_{3}(\mathbf{d}) = \sqrt
{f_2(\mathbf{d})^2 + f_5(\mathbf{d})^2} \\
\text{Subject to:} \quad & g_i(\mathbf{d}) \leq 0 \\
& h_k(\mathbf{d}) = 0 \\
& d_{i,L} \leq d_i \leq d_{i,u}, ; i=1, ..., N_d
\end{aligned}
\end{equation}

\item
  \begin{quote}
  \(F_{4}\): Minimization of the gear mass, \(H_{\text{VL}}\) and
  differences of the specific sliding at the roots and maximization of
  \(S_{H}\), \(S_{F}\) and contact ratio.
  \end{quote}

 \begin{equation}
    \begin{aligned}
\text{Minimize:} \quad & F_{4}(\mathbf{d}) = \sqrt{\sum_{i=2,5,6} f_i(\mathbf{d})^2 + \sum_{j=1,3,4} \left( \frac{1}{1 + f_j(\mathbf{d})} \right)^2} \\
\text{Subject to:} \quad & g_i(\mathbf{d}) \leq 0 \\
& h_k(\mathbf{d}) = 0 \\
& d_{i,L} \leq d_i \leq d_{i,u}, \; i=1, ..., N_d
\end{aligned}
\end{equation}

\end{itemize}
\item
  Study 3. Variation of the PSO configurations: using the mass
  minimization as the objective and tolerance
  \(\text{to}l_{u}^{M} = 0.2\), the following variations have been
  considered so that the computational time is equal for all of the
  variations; therefore, a fair comparison between them is possible:

  \begin{itemize}
  \item
    \begin{quote}
    M1: RPSO with a population of 10 particles (\#POP = 10) and 10000
    generations
    \end{quote}
  \item
    \begin{quote}
    M2: PSO with a population of 10 particles (\#POP = 10) and 10000
    generations
    \end{quote}
  \item
    \begin{quote}
    M3: RPSO with a population of 50 particles (\#POP = 50) and 2000
    generations
    \end{quote}
  \item
    \begin{quote}
    M4: PSO with a population of 50 particles (\#POP = 50) and 2000
    generations
    \end{quote}
  \item
    \begin{quote}
    M5: RPSO with a population of 500 particles (\#POP = 500) and 200
    generations
    \end{quote}
  \item
    \begin{quote}
    M6: PSO with a population of 500 particles (\#POP = 500) and 200
    generations
    \end{quote}
  \end{itemize}
\end{itemize}

\begin{enumerate}
\def\labelenumi{\arabic{enumi}.}
\setcounter{enumi}{3}
\item
  \textbf{\uline{Results and discussion}}
\end{enumerate}

The results of spur gear optimization are documented in this section.
For reasons related to the comparison with literature studies, two case
studies are driven and used to execute the algorithm and to obtain the
results. The dependent variables related to the center distance,
transmission ratio, lubricant, material properties, input power, and
speed of the pinion are listed in Table \ref{tab:2}, according to the selected
gear from each literature study. Properties such as input power,
transmission ratio, and center distance, which are significant inputs of
the proposed algorithm, were calculated based on the gear variables of Table \ref{tab:3}.

The conducted research on gear optimization has played an important
role in the exposition of a large set of results from different
numerical experiments. The existent commercial gear tooth contact
analysis software \emph{KISSsoft} was used for validating the results of
the model implemented in the present study.

\begin{longtable}{ccc p{10cm}}
    \caption{General and specific gear properties considered in this work
    for comparison} \label{tab:2} \\
    \toprule
    \multicolumn{4}{c}{General Properties} \\
    \midrule
    \multicolumn{4}{p{15cm}}{Oil: PAO ISO VG 150; Material: 20MnCr5; Material Young Modulus: 210 GPa; Material Poisson ratio: 0.3; \(T_{\text{lub}} = 70{^\circ}C\)} \\
    \midrule
    Ref. & Authors & Abbr. & Specific properties \\
    \midrule
    \endfirsthead

[\ref{ref:9}] & Miler \emph{et al.} & C1 & Material specific gravity: 7.84;
Input power: 10.053 kW; Speed of the pinion: 960 rpm; Center distance:
199.845 mm; Transmission ratio: \(u = 3.55\) \\

[\ref{ref:5}] & Maputi \emph{et al.} & C2 & Material specific gravity: 8;
Input power: 12.480 kW; Speed of the pinion: 1500 rpm; Center distance:
130.625 mm; Transmission ratio: \(u = 4\) \\
\bottomrule
\end{longtable}

\begin{longtable}{cccc}
\caption{Selected gears from two literature references} \label{tab:3} \\
\toprule
\multicolumn{2}{c}{Variable} & C1 & C2 \\
\midrule
\endfirsthead
\multirow{7}{*}{Design variables} & \(\alpha\) {[}º{]} & 20 & 25 \\
& \(z_{1}\) & 23 & 19 \\
& \(z_{2}\) & 82 & 76 \\
& \(x_{1}\) & 0.699 & 0 \\
& \(x_{2}\) & 0.136 & 0 \\
& \(m\) {[}mm{]} & 3.75 & 2.75 \\
& \(b\) {[}mm{]} & 22.5 & 34.84 \\
\midrule
\multirow{6}{*}{Objective functions} & \(\epsilon_{\alpha}\) & 1.458 &
1.488 \\
& \(H_{\text{VL}}\) & 0.141 & 0.129 \\
& \(\Delta g\) & 0.628 & 0.870 \\
& Mass {[}kg{]} & 14.347 & 10.160 \\
& \(S_{H}\) & 2.13 & 1.69 \\
& \(S_{F}\) & 7.65 & 6.93 \\
\bottomrule
\end{longtable}

In Table \ref{tab:4}, results on the minimization of the gear mass are
represented. Several conclusions can be taken from it. Firstly, the mass
has been lower than the literature for the studied cases [\ref{ref:5}],
[\ref{ref:9}]. This reduction corresponds to from 51\% to 55\%. The reduction
of the masses had been previously hypothesized since both safety factors
of the gears of each of the original gears were not near the allowable
values. With the decrease of the module, the minimum face width value of
10 mm has been achieved for case C1. In C2, the face width value is not
decreased to its minimum since the limit of the safety factor \(S_{H}\)
has been reached.

Moreover, it is possible to verify that no significant difference
between tolerance values exists. However, a tolerance of 0.2 is
preferred since it takes a lower computational time. With the decrease
of the mass, \(H_{\text{VL}}\) and the difference in the maximum specific
sliding have also been reduced. The contact ratio is increased.

\begin{longtable}{@{} l *{6}{S[table-format=-1.4]} @{}}
\caption{Optimized gears for each of the case studies C\textsubscript{k}
for different tolerances.} \label{tab:4} \\
\toprule
\multicolumn{2}{c}{Variable} & \multicolumn{2}{c}{C1 (\(F_{1}\))} & \multicolumn{2}{c}{C2 (\(F_{1}\))} \\
\cmidrule(lr){3-4} \cmidrule(lr){5-6}
& & {\(tol_{u}^{M} = 0.05\)} & {\(tol_{u}^{M} = 0.2\)} & {\(tol_{u}^{M} = 0.05\)} & {\(tol_{u}^{M} = 0.2\)} \\
\midrule
\endfirsthead
\multicolumn{5}{c}%
{{\bfseries Table \thetable\ continued from previous page}} \\
\toprule
& \multicolumn{2}{c}{C1 (\(F_{1}\))} & \multicolumn{2}{c}{C2 (\(F_{2}\))} \\
\cmidrule(lr){2-3} \cmidrule(lr){4-5}
Variable & {\(tol_{u}^{M} = 0.05\)} & {\(tol_{u}^{M} = 0.2\)} & {\(tol_{u}^{M} = 0.05\)} & {\(tol_{u}^{M} = 0.2\)} \\
\midrule
\endhead
\multirow{7}{*}{\parbox{3cm}{Design variables}} & \(\alpha\) {[}º{]} & 20 & 20 & 25 &
25 \\
& \(z_{1}\) & 44 & 44 & 26 & 26 \\
& \(z_{2}\) & 156 & 156 & 104 & 104 \\
& \(x_{1}\) & 0.3175 & 0.1917 & 0.4263 & 0.4259 \\
& \(x_{2}\) & -0.3946 & -0.2688 & -0.1104 & -0.1100 \\
& \(m\) {[}mm{]} & 2 & 2 & 2 & 2 \\
& \(b\) {[}mm{]} & 10 & 10 & 16.999 & 16.988 \\
\midrule
\multirow{6}{*}{\parbox{3cm}{Output variables / objective functions}} &
\(\epsilon_{\alpha}\) & 1.764 & 1.788 & 1.445 & 1.445 \\
& \(H_{\text{VL}}\) & 0.082 & 0.077 & 0.105 & 0.105 \\
& \(\Delta g\) & 0.213 & 0.038 & 0.246 & 0.246 \\
& \(S_{H}\) & 1.56 & 1.58 & 1.40 & 1.40 \\
& \(S_{F}\) & 2.39 & 2.41 & 2.96 & 2.96 \\
& \textbf{\uline{Mass {[}kg{]}}} & \textbf{\uline{6.424}} &
\textbf{\uline{6.438}} & \textbf{\uline{4.909}} &
\textbf{\uline{4.906}} \\
\bottomrule
\end{longtable}

In Table \ref{tab:5}, the results of the second study are shown. For the objective
function \(F_{2}\), it has been possible to decrease the mass by 3\% and
25\% while still increasing the safety factor \(S_{F}\). Module values
and face width are now higher, so the root stress is decreased and the
safety factor \(S_{F}\) is increased. For the objective function
\(F_{3}\), the mass has been significantly reduced, as occurred in the
situation where mass reduction is the single objective. The
\(H_{\text{VL}}\) decreases as the module decreases and the number of
teeth increases. This reduction in the module and the increase of the
number of teeth promotes a possible reduction in width and,
consequently, in the gear mass. The module and face width values are
similar to the single objective.

Moreover, reductions of 80\% and 98\% on the value of \(\Delta g\), when
comparing to the literature [\ref{ref:5}], [\ref{ref:9}], are observed even if it
was not aimed to minimize it. When it comes to trying to minimize and
maximizing all the output variables of interest, using the objective
function \(F_{4}\), one can observe that some objectives are indeed
improved, but, as a consequence, the others have been penalized. For
instance, in C1, a reduction of 94\% of the \(\Delta g\), an increase in
16\% of the contact ratio and a reduction of 26\% of the
\(H_{\text{VL}}\) are obtained, but the mass is now higher, and safety
factors are lower than the original gear. In C2, mass and \(\Delta g\)
are improved, but the same does not happen with the other objectives.

\begin{longtable}{@{} l *{8}{S[table-format=-1.4]} @{}}
\caption{Optimized gears for each of the case studies C\textsubscript{k}
for different multi-objective problems.} \label{tab:5} \\
\toprule
\multicolumn{2}{c}{Variable} & \multicolumn{3}{c}{C1 (\(tol_{u}^{M}=0-2\))} & \multicolumn{3}{c}{C2 (\(tol_{u}^{M} = 0.2\))} \\
\cmidrule(lr){3-5} \cmidrule(lr){6-8}
& & {\(F_2\)} & {\(F_3\)} & {\(F_4\)} & {\(F_2\)} & {\(F_3\)} & {\(F_4\)} \\
\midrule
\endfirsthead
\multirow{7}{*}{\parbox{3cm}{Design variables}} & \(\alpha\) {[}º{]} & 20 & 20 & 20 & 25 &
25 & 25 \\
& \(z_{1}\) & 22 & 44 & 30 & 18 & 26 & 18 \\
& \(z_{2}\) & 78 & 156 & 103 & 69 & 104 & 70 \\
& \(x_{1}\) & 0.4506 & 0.0823 & 0.2509 & 0.1725 & 0.2682 & 0.0528  \\
& \(x_{2}\) & -0.4891 & -0.1594 & -0.1350 & -0.1308 & 0.0477 & -0.4998 \\
& \(m\) {[}mm{]} & 4 & 2 & 3 & 3 & 2 & 3 \\
& \(b\) {[}mm{]} & 21.866 & 10 & 25.052 & 26.517 & 17.697 & 26.095 \\
\midrule
\multirow{6}{*}{\parbox{3cm}{Output variables / objective functions}} &
\(\epsilon_{\alpha}\) & 1.598 & 1.807 & \textbf{\uline{1.696}} & 1.449 & 1.467 & \textbf{\uline{1.506}} \\
& \(H_{\text{VL}}\) & 0.153 & \textbf{\uline{0.076}} & \textbf{\uline{0.105}} & 0.134 & \textbf{\uline{0.096}} & \textbf{\uline{0.139}} \\
& \(\Delta g\) & 0.575 & 0.126 & \textbf{\uline{0.041}} & 0.333 & 0.018 & \textbf{\uline{0.649}} \\
& Mass {[}kg{]} & \textbf{\uline{13.909}} & \textbf{\uline{6.450}} & \textbf{\uline{15.944}} & \textbf{\uline{7.590}} & \textbf{\uline{5.133}} &  \textbf{\uline{7.509}} \\
& \(S_{H}\) & \textbf{\uline{2.07}} & 1.57 & \textbf{\uline{2.11}} & \textbf{\uline{1.59}} & 1.40 & \textbf{\uline{1.56}} \\
& \(S_{F}\) & \textbf{\uline{8.55}} & 2.42 &
\textbf{\uline{6.69}} & \textbf{\uline{6.31}} & 3.07 & 
\textbf{\uline{6.40}} \\
\bottomrule
\end{longtable}

\begin{figure}[ht]
\centering
\includegraphics[width=0.85\linewidth]{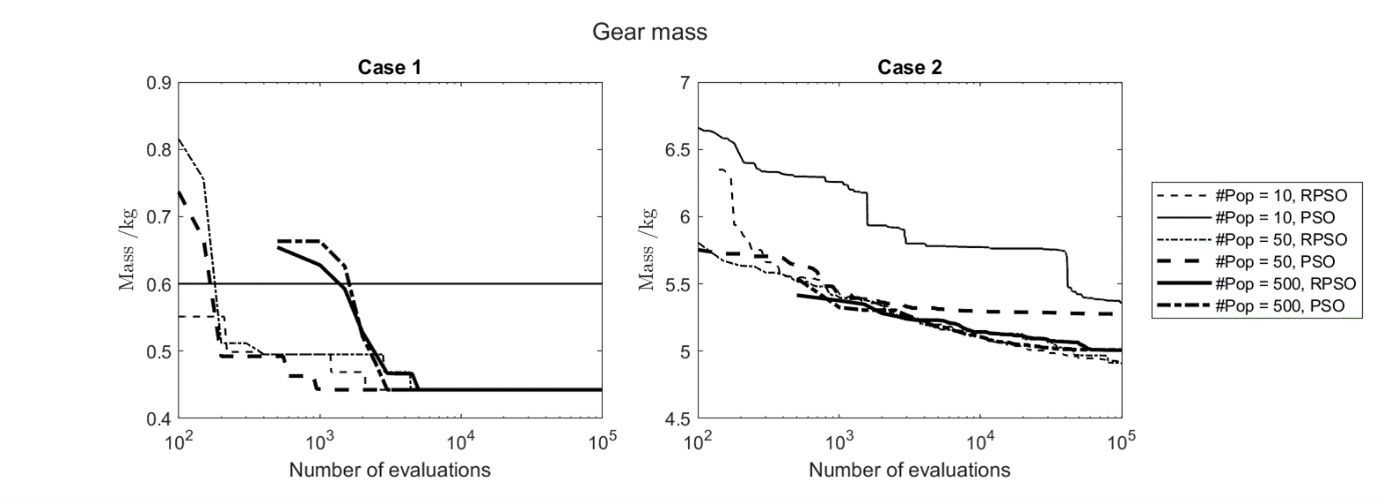}
\caption{Gear mass variation along the number of evaluations.}
\label{fig:3}
\end{figure}

Lastly, the mass reduction optimization results for different selected
configurations of PSO variations (of population size and random search)
have been obtained. Since the resulting gear of the optimization process
for the mass reduction is already known, the objective of this study is
to observe the behavior of the different configurations along with the
iteration and the selection of the most suitable for this problem. The
plot corresponding to the average (on ten simulation repetitions) of the
minimum gear mass for each case and generation is represented in Figure \ref{fig:3}.

\begin{figure}[ht]
\centering
\includegraphics[width=\linewidth]{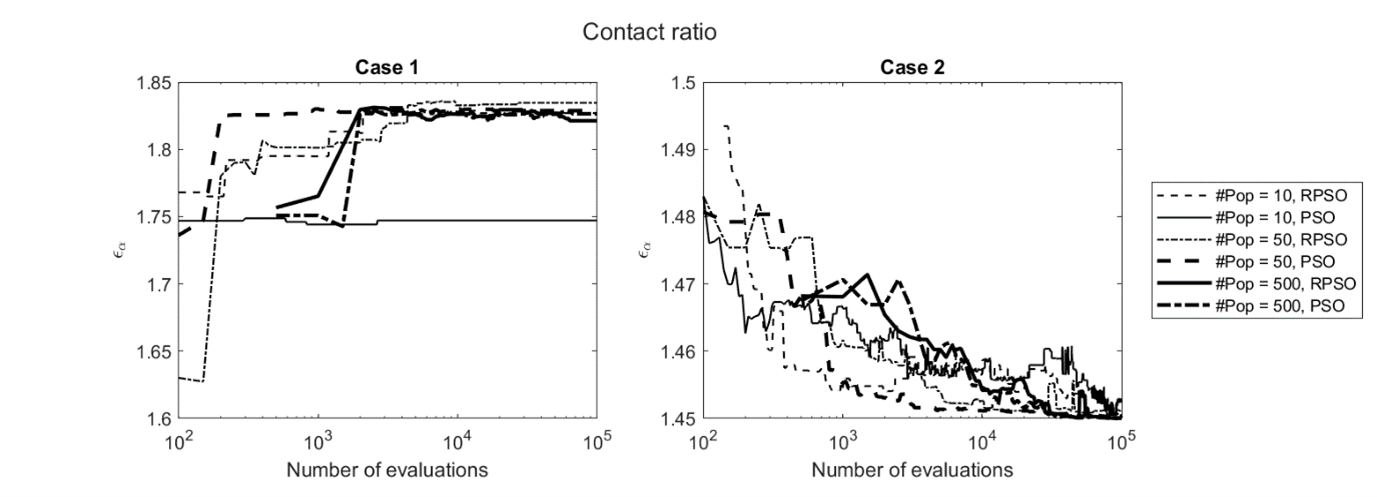}
\caption{Gear contact ratio variation along with the number of
evaluations. In Case 1, the contact ratio is higher than 1.8 if the method is well chosen; in Case 2, the contact ratio tends to the
constrained value of 1.45.}
\label{fig:4}
\end{figure}

According to Figure \ref{fig:4}, in case C1, the gear mass is reduced and
stabilized in less than 10000 iterations for most configurations. When
using the PSO local search with ten particles, the difficulty of finding
the global minimum is evident. However, for case C2, no gear mass
stabilization is apparent, even for 100000 iterations. Using PSO with 10
particles continue to be not suitable for the problem since the average
gear mass is always higher than the rest of the configurations for any
of the number of generations stopping criteria. However, using RPSO
configurations is the most suitable option, especially for a smaller
number of particles, resulting from the complexity of the problem of
gear mass reduction.

Moreover, the figures in Appendix A (5 to 14) have resulted from the
variation of other relevant gear variables with the iterations.
Concerning both safety factors, for both case studies, a decrease is
observed. As previously observed, when minimizing the mass, the safety
factor \(S_{H}\) is also minimized until its convergence to the
constraint. Also, the minimum specific sliding and \(H_{\text{VL}}\) are
decreased in general. However, the contact ratio may decrease or
increase depending on the case study. In C1, the evident increase in the
contact ratio is observed, and its convergence happens around no more
than 1.85. In opposite, the contact ratio decreases in C2, in which the
value stabilized around the constraint value of 1.45.

Concerning the design variables of the gear, it is possible to observe
that the number of teeth of the pinion has a slight tendency to
increase. In contrast, the module tends to decrease to satisfy the
gear's required center distance. On average, the module's value is never
no more than 2 mm. Moreover, in C2, this value is only achieved at the
end of several iterations. The face width is also decreasing at the end
of the iterations, and the pinion and wheel shift coefficients in C1.
These coefficients tend to increase in C2.

\begin{enumerate}
\def\labelenumi{\arabic{enumi}.}
\setcounter{enumi}{4}
\item
  \textbf{Conclusions}
\end{enumerate}

The present work is aimed at solving the design problem of spur gears
under the usual transmission ratio and center distance requirements. The
optimized designs have been compared to selected gears from the recent
literature. Single and multiple-objective problems have been considered:
1. gear mass reduction; 2. gear mass reduction and gear loss factor
\(H_{\text{VL}}\); 3. gear mass reduction and increase of safety factors
\(S_{F}\) and \(S_{H}\); 4. gear mass, specific sliding and
\(H_{\text{VL}}\) reductions and increase of \(S_{F}\), \(S_{H}\), and
contact ratio. The main geometrical parameters of the spur gear are the
design variables: pinion and wheel number of teeth and profile shift
coefficients, module and face width. There are constraints considered in
the present work: \(S_{H}\) and \(S_{F}\) may not be lower than 1.4 and
2, respectively, and the contact ratio may not be lower than 1.4. Also,
the face width is considered between 6 and 12 times the value of the
module. Moreover, module values are standard. The main results are as
follows:

\begin{itemize}
\item
  Gear mass reductions of 51\% to 55\% have been achieved compared to
  the literature [\ref{ref:5}], [\ref{ref:9}] due to the reduction of the module's
  value and, consequently, the face width.
\item
  The reduction of the gear mass led to safety factors near their
  possible minima.
\item
  The difference between specific sliding decreases by up to 98\%,
  compared to the literature [\ref{ref:5}], [\ref{ref:9}], in multi-objective
  optimization.
\item
  The gear mass increases when more objectives are considered.
\end{itemize}

However, further investigations need to be carried out concerning the
optimization problems of spur gears. Firstly, solving the optimization
problem is costly, and the considered feasible search space is
significantly big. Therefore, a different approach regarding reducing
the feasible search space to a search space near the safety factor
constraints can be considered in the future. Also, the significant
amount of mechanical and geometrical constraints regarding the profile
shift coefficients may have been driven to an almost random search; the
PSO algorithm controls it if other design variables such as the number
of teeth and the module do not change. Therefore, the proposed algorithm
may also be a topic of interest for further improvements in future
works.

The significant contribution of this work is an optimization algorithm
that proposes a suitable spur gear for specific objectives and with
requirements of transmission ratio and center distance. In the future,
the development of an interface for choosing the optimized spur gear
based on objectives that the user weights would be a topic of interest
for the scientific community's support.

\vspace{4cm}
\textbf{\uline{References:}}

\begin{enumerate}
\item \label{ref:1} E. Maier, A. Ziegltrum, T. Lohner, and K. Stahl,
`Characterization of TEHL contacts of thermoplastic gears',
\emph{Forsch. im Ingenieurwes.}, vol. 81, no. 2, pp. 317--324, 2017,
doi: 10.1007/s10010-017-0230-4.

\item \label{ref:2} D. J. Politis, N. J. Politis, and J. Lin, `Review of recent
developments in manufacturing lightweight multi-metal gears',
\emph{Prod. Eng.}, vol. 15, no. 2, pp. 235--262, 2021, doi:
10.1007/s11740-020-01011-5.

\item \label{ref:3} Q. Yao, `Multi-objective optimization design of spur gear based
on NSGA-II and decision making', \emph{Adv. Mech. Eng.}, vol. 11, no. 3,
pp. 1--8, 2019, doi: 10.1177/1687814018824936.

\item \label{ref:4} S. Li and A. Kahraman, `A scuffing model for spur gear
contacts', \emph{Mech. Mach. Theory}, vol. 156, p. 104161, 2021, doi:
https://doi.org/10.1016/j.mechmachtheory.2020.104161.

\item \label{ref:5} E. S. Maputi and R. Arora, `Multi-objective spur gear design
using teaching learning-based optimization and decision-making
techniques', \emph{Cogent Eng.}, vol. 6, no. 1, p. 1665396, Jan. 2019,
doi: 10.1080/23311916.2019.1665396.

\item \label{ref:6} Maputi, Edmund S. and Arora, Rajesh, `Design optimization of a
three-stage transmission using advanced optimization techniques',
\emph{Int. J. Simul. Multidisci. Des. Optim.}, vol. 10, p. A8, 2019,
doi: 10.1051/smdo/2019009.

\item \label{ref:7} Z. Chen, Y. Jiang, Z. Tong, and S. Tong, `Residual Stress
Distribution Design for Gear Surfaces Based on Genetic Algorithm
Optimization', \emph{Materials} , vol. 14, no. 2. 2021, doi:
10.3390/ma14020366.

\item \label{ref:8} D. Miler, D. Žeželj, A. Lončar, and K. Vučković,
`Multi-objective spur gear pair optimization focused on volume and
efficiency', \emph{Mech. Mach. Theory}, vol. 125, pp. 185--195, 2018,
doi: https://doi.org/10.1016/j.mechmachtheory.2018.03.012.

\item \label{ref:9} D. Miler, A. Lončar, D. Žeželj, and Z. Domitran, `Influence of
profile shift on the spur gear pair optimization', \emph{Mech. Mach.
Theory}, vol. 117, pp. 189--197, 2017, doi:
https://doi.org/10.1016/j.mechmachtheory.2017.07.001.

\item \label{ref:10} H. Abderazek, D. Ferhat, I. Atanasovska, and K. Boualem, `A
differential evolution algorithm for tooth profile optimization with
respect to balancing specific sliding coefficients of involute
cylindrical spur and helical gears', \emph{Adv. Mech. Eng.}, vol. 7, no.
9, pp. 1--11, 2015, doi: 10.1177/1687814015605008.

\item \label{ref:11} J. A. Korta and D. Mundo, `A population-based meta-heuristic
approach for robust micro-geometry optimization of tooth profile in spur
gears considering manufacturing uncertainties', \emph{Meccanica}, vol.
53, no. 1, pp. 447--464, 2018, doi: 10.1007/s11012-017-0737-7.

\item \label{ref:12} K. Daoudi and E. M. Boudi, `Genetic Algorithm Approach for Spur
Gears Design Optimization', in \emph{2018 International Conference on
Electronics, Control, Optimization and Computer Science (ICECOCS)},
2018, pp. 1--5, doi: 10.1109/ICECOCS.2018.8610520.

\item \label{ref:13} M. Hofstetter, D. Lechleitner, M. Hirz, M. Gintzel, and A.
Schmidhofer, `Multi-objective gearbox design optimization for xEV-axle
drives under consideration of package restrictions', \emph{Forsch. im
Ingenieurwesen/Engineering Res.}, vol. 82, no. 4, pp. 361--370, 2018,
doi: 10.1007/s10010-018-0278-9.

\item \label{ref:14} K. Deb, S. Agrawal, A. Pratap, and T. Meyarivan, `A Fast
Elitist Non-dominated Sorting Genetic Algorithm for Multi-objective
Optimization: NSGA-II BT - Parallel Problem Solving from Nature PPSN
VI', 2000, pp. 849--858.

\item \label{ref:15} Ü. Atila, M. Dörterler, R. Durgut, and İ. Şahin, `A
comprehensive investigation into the performance of optimization methods
in spur gear design', \emph{Eng. Optim.}, vol. 52, no. 6, pp.
1052--1067, Jun. 2020, doi: 10.1080/0305215X.2019.1634702.

\item \label{ref:16} T. Yokota, T. Taguchi, and M. Gen, `A solution method for
optimal weight design problem of the gear using genetic algorithms',
\emph{Comput. Ind. Eng.}, vol. 35, no. 3, pp. 523--526, 1998, doi:
https://doi.org/10.1016/S0360-8352(98)00149-1.

\item \label{ref:17} V. Savsani, R. V Rao, and D. P. Vakharia, `Optimal weight
design of a gear train using particle swarm optimization and simulated
annealing algorithms', \emph{Mech. Mach. Theory}, vol. 45, no. 3, pp.
531--541, 2010, doi:
https://doi.org/10.1016/j.mechmachtheory.2009.10.010.

\item \label{ref:18} N. Godwin Raja Ebenezer, S. Ramabalan, and S.
Navaneethasanthakumar, `Design optimisation of mating helical gears with
profile shift using nature inspired algorithms', \emph{Aust. J. Mech.
Eng.}, pp. 1--8, May 2020, doi: 10.1080/14484846.2020.1761007.

\item \label{ref:19} R. Guilbault and S. Lalonde, `Tip relief designed to optimize
contact fatigue life of spur gears using adapted PSO and Firefly
algorithms', \emph{SN Appl. Sci.}, vol. 3, no. 1, p. 66, 2021, doi:
10.1007/s42452-020-04129-4.

\item \label{ref:20} Z. Tomori, `An Optimal Choice of Profile Shift Coefficients for
Spur Gears', \emph{Machines} , vol. 9, no. 6. 2021, doi:
10.3390/machines9060106.

\item \label{ref:21} J. Tavčar, B. Černe, J. Duhovnik, and D. Zorko, `A
multicriteria function for polymer gear design optimization', \emph{J.
Comput. Des. Eng.}, vol. 8, no. 2, pp. 581--599, 2021, doi:
10.1093/jcde/qwaa097.

\item \label{ref:22} P. Rai and A. G. Barman, `Tooth Profile Optimization of Helical
Gear with Balanced Specific Sliding Using TLBO Algorithm BT - Advanced
Engineering Optimization Through Intelligent Techniques', 2020, pp.
203--210.

\item \label{ref:23} J. Wang and I. Howard, `Finite element analysis of High Contact
Ratio spur gears in mesh', \emph{J. Tribol.}, vol. 127, no. 3, pp.
469--483, 2005, doi: 10.1115/1.1843154.

\item \label{ref:24} N. Gunantara, `A review of multi-objective optimization:
Methods and its applications', \emph{Cogent Eng.}, vol. 5, no. 1, p.
1502242, Jan. 2018, doi: 10.1080/23311916.2018.1502242.

\item \label{ref:25} C. António, \emph{Otimização de Sistemas em Engenharia:
Fundamentos e algoritmos para o projeto ótimo}, Quântica E. Porto, 2020.

\item \label{ref:26} Y. Sun, Y. Gao, and X. Shi, `Chaotic multi-objective particle
swarm optimization algorithm incorporating clone immunity',
\emph{Mathematics}, vol. 7, no. 2, pp. 1--16, 2019, doi:
10.3390/math7020146 M4 - Citavi.

\item \label{ref:27} R. Eberhart and J. Kennedy, `A new optimizer using particle
swarm theory', in \emph{MHS'95. Proceedings of the Sixth International
Symposium on Micro Machine and Human Science}, Oct. 1995, pp. 39--43,
doi: 10.1109/MHS.1995.494215.

\item \label{ref:28} D. ping Tian, `A Review of Convergence Analysis of Particle
Swarm Optimization', \emph{Int. J. Grid Distrib. Comput.}, vol. 6, no.
6, pp. 117--128, 2013, doi: 10.14257/ijgdc.2013.6.6.10 M4 - Citavi.

\item \label{ref:29} Y. Shi and R. Eberhart, `A modified particle swarm optimizer',
in \emph{1998 IEEE International Conference on Evolutionary Computation
Proceedings. IEEE World Congress on Computational Intelligence (Cat.
No.98TH8360)}, May 1998, pp. 69--73, doi: 10.1109/ICEC.1998.699146.

\item \label{ref:30} M. Clerc, `The swarm and the queen: towards a deterministic and
adaptive particle swarm optimization', in \emph{Proceedings of the 1999
Congress on Evolutionary Computation-CEC99 (Cat. No. 99TH8406)}, Jul.
1999, vol. 3, pp. 1951-1957 Vol. 3, doi: 10.1109/CEC.1999.785513.

\item \label{ref:31} M. Clerc and J. Kennedy, `The particle swarm - explosion,
stability, and convergence in a multidimensional complex space',
\emph{IEEE Trans. Evol. Comput.}, vol. 6, no. 1, pp. 58--73, Feb. 2002,
doi: 10.1109/4235.985692.

\item \label{ref:32} F. van den Bergh, `An Analysis of Particle Swarm Optimizers',
no. November, p. 315, 2001.

\item \label{ref:33} R. Fitas, `Optimal Design of Composite Structures using the
Particle Swarm Method and Hybridizations', University of Porto, 2022.

\item \label{ref:34} C. Fitas, Ricardo; Hesseler, Stefan; Wist, Santino; Greb,
`Kinematic Draping Simulation Optimization of a Composite B-Pillar
Geometry using Particle Swarm Optimization'.

\item \label{ref:35} R. Fitas, G. das Neves Carneiro, and C. Conceição António, `An elitist multi-objective particle swarm optimization algorithm for
composite structures design', \emph{Compos. Struct.}, vol. 300, p.
116158, 2022, doi: https://doi.org/10.1016/j.compstruct.2022.116158.
\end{enumerate}

\vspace{2cm}
\hypertarget{annex-a}{%
\subsection{Annex A}\label{annex-a}}

\begin{figure}[ht]
\centering
\includegraphics[width=\linewidth]{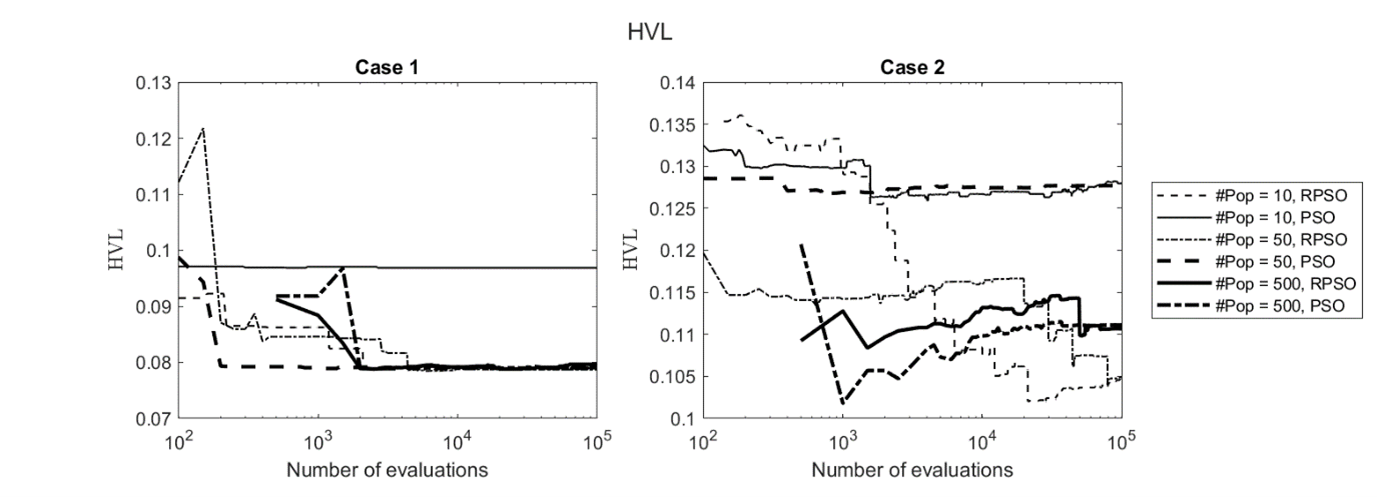}
\caption{Gear loss factor variation along with the number of
evaluations. In general, \(H_{\text{VL}}\) tends to have a lower value
for a reduced mass.}
\label{fig:5}
\end{figure}

\begin{figure}[ht]
\centering
\includegraphics[width=\linewidth]{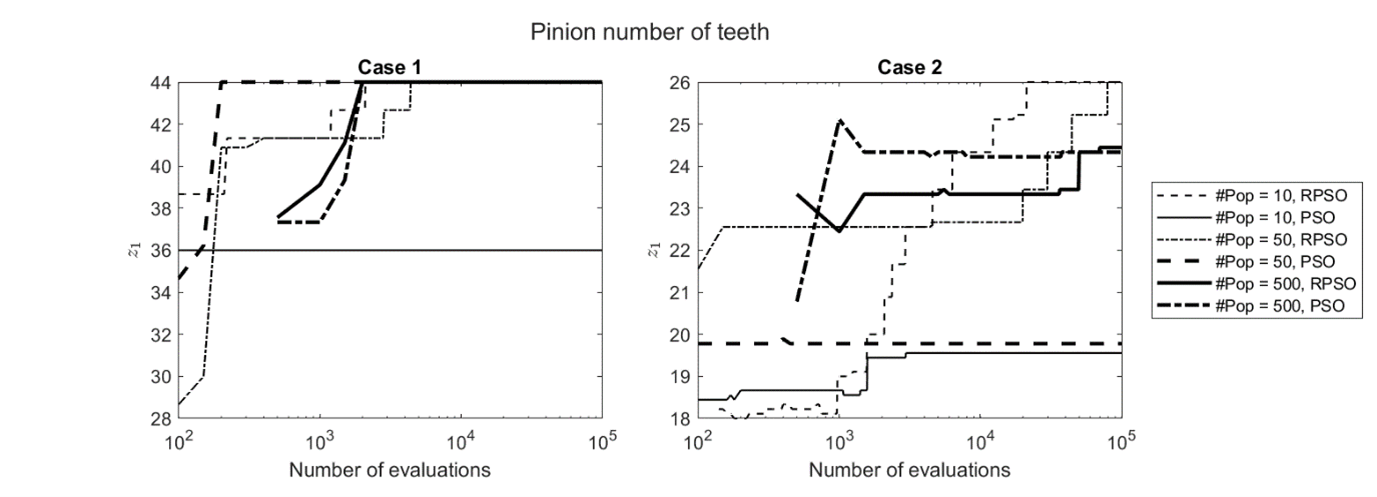}
\caption{Variation of the pinion number of teeth along with the number
of evaluations. In general, the number of teeth in the pinion tends to
increase to achieve a reduced gear mass.}
\label{fig:6}
\end{figure}

\begin{figure}[ht]
\centering
\includegraphics[width=\linewidth]{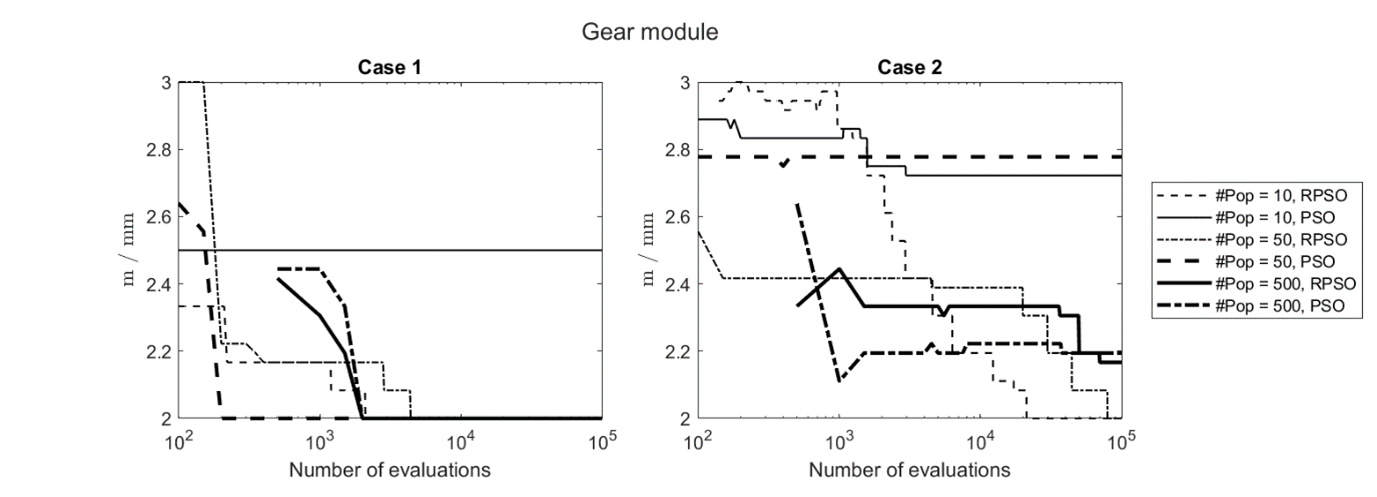}
\caption{Gear module variation along with the number of evaluations. In
general, the module tends to decrease to achieve a reduced gear mass.}
\label{fig:7}
\end{figure}

\begin{figure}[ht]
\centering
\includegraphics[width=\linewidth]{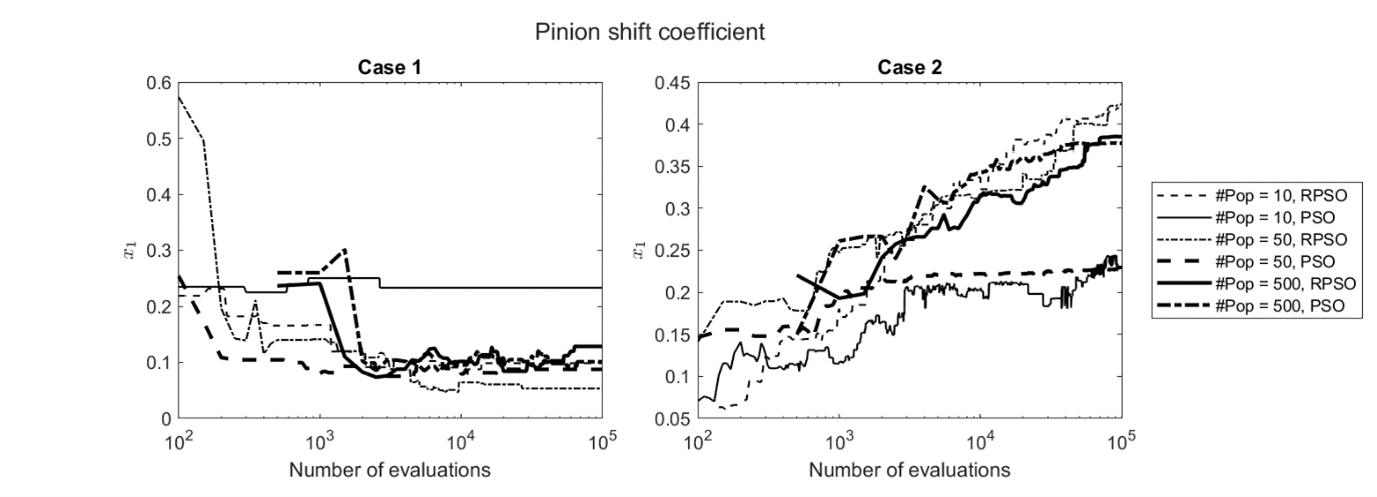}
\caption{Pinion shift coefficient variation along with the number of
evaluations. In Case 1, the shift coefficient tends to have a lower
value; in Case 2, the value is increased.}
\label{fig:8}
\end{figure}

\begin{figure}[ht]
\centering
\includegraphics[width=\linewidth]{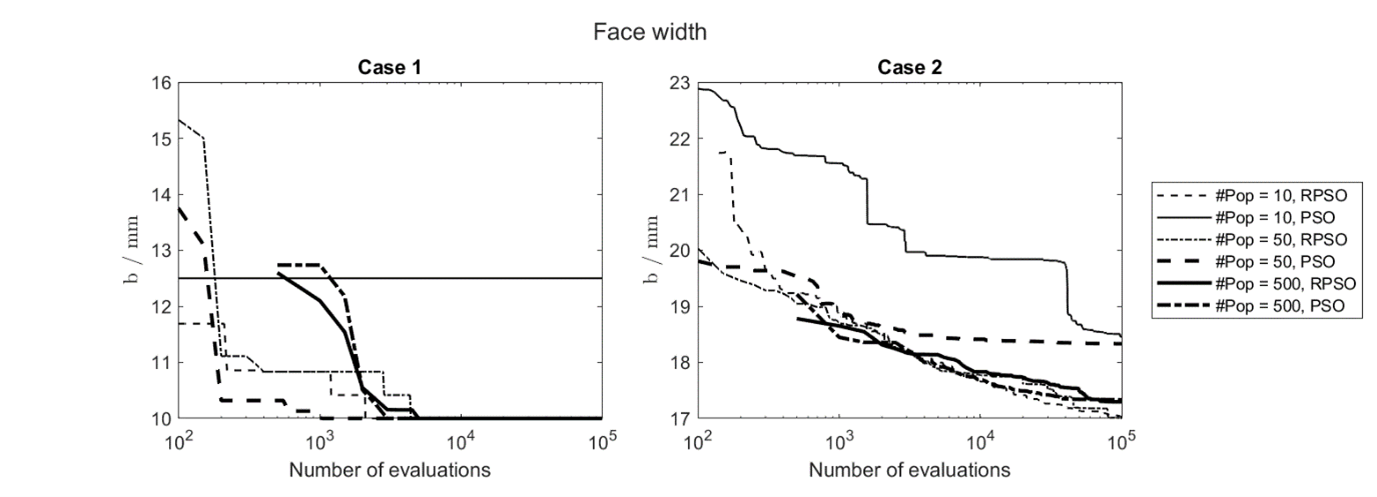}
\caption{Gear face width variation along with the number of
evaluations. In general, the face width is reduced for the achievement
of a reduced gear mass.}
\label{fig:9}
\end{figure}

\begin{figure}[ht]
\centering
\includegraphics[width=\linewidth]{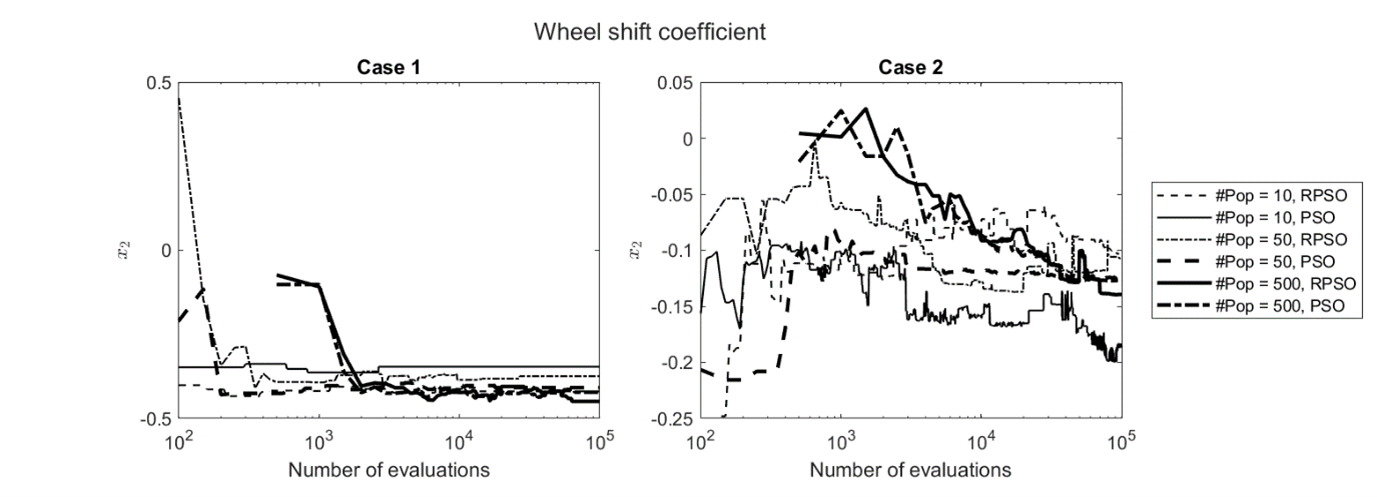}
\caption{Wheel shift coefficient variation along with the number of
evaluations. In Case 1, this shift coefficient tends to have a minimum value corresponding to the side constraint of the problem; in Case 2,
the coefficient tends to decrease.}
\label{fig:10}
\end{figure}

\begin{figure}[ht]
\centering
\includegraphics[width=\linewidth]{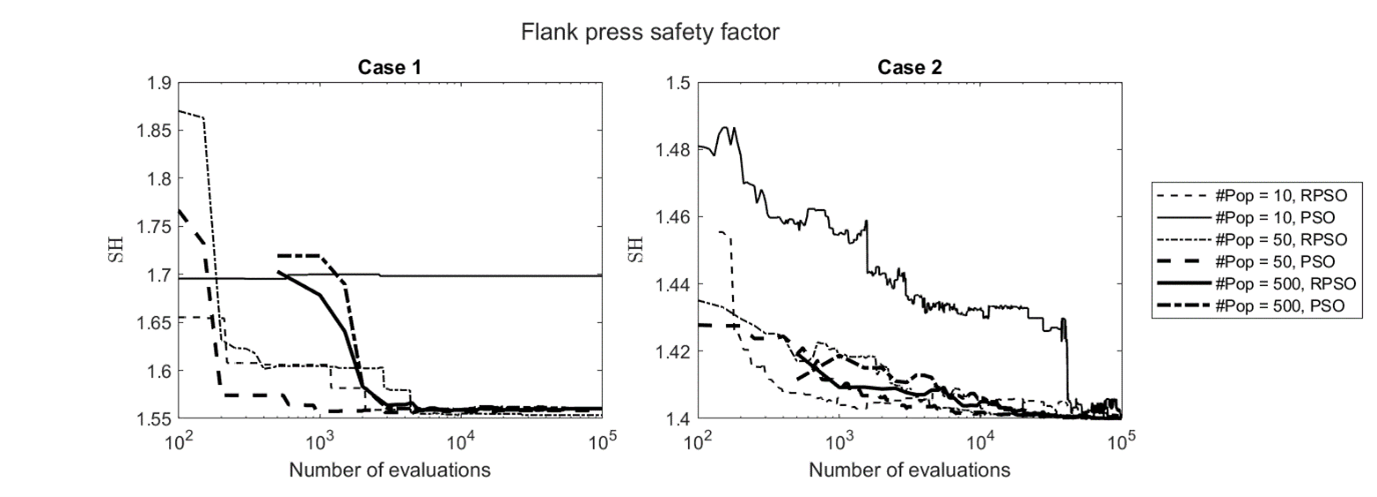}
\caption{Flank press safety factor variation along with the number of
evaluations. In general, since the face width decreases, the pressure
increases, reducing the safety factor.}
\label{fig:11}
\end{figure}

\begin{figure}[ht]
\centering
\includegraphics[width=\linewidth]{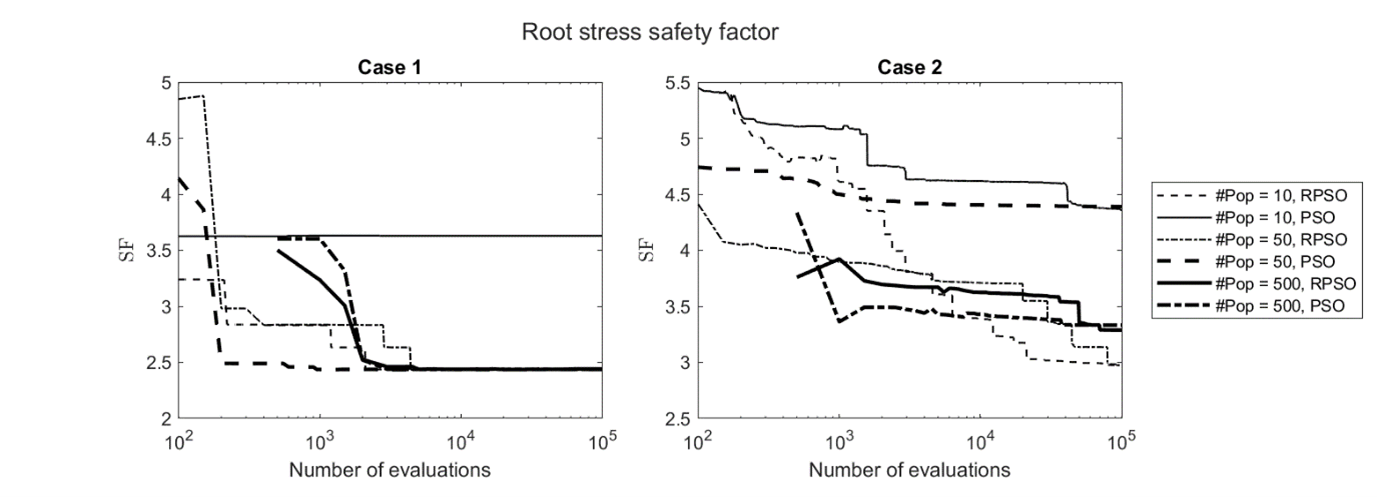}
\caption{Root stress safety factor variation along with the number of
evaluations. In general, since the face width decreases, the pressure
increases, reducing the safety factor.}
\label{fig:12}
\end{figure}

\begin{figure}[ht]
\centering
\includegraphics[width=\linewidth]{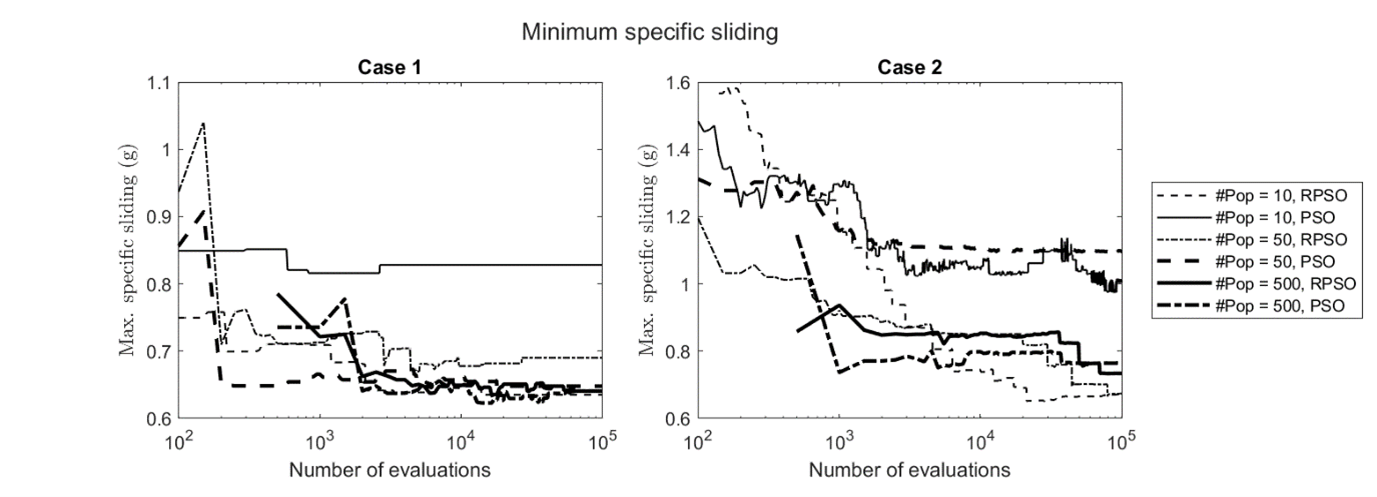}
\caption{Specific sliding variation along with the number of
evaluations. In general, the specific sliding decreases with the
decrease of the gear mass.}
\label{fig:13}
\end{figure}

\end{document}